# Magnetic field dependence of low temperature specific heat of spinel oxide superconductor $LiTi_2O_4$

C. P. Sun,[a] J. -Y. Lin,[b] S. Mollah,[a,†] P. L. Ho,[a] H. D. Yang,[a,*] F. C. Hsu,[c] Y. C. Liao,[d] and M. K. Wu[e,f]

[a]*Department of Physics, National Sun Yat Sen University, Kaohsiung 804, Taiwan, ROC.*
[b]*Institute of Physics, National Chiao-Tung University, Hsin-chu 300, Taiwan, ROC.*
[c]*Department of Materials Science and Engineering, National Tsing Hua University, Hsinchu, Taiwan, ROC*
[d]*Department of Physics and Materials Science Center, National Tsing Hua University, Hsinchu, Taiwan, ROC*
[e]*Department of Physics, National Tsing Hwa University, Hsin-chu 300, Taiwan, ROC.*
[f]*Institute of Physics, Academia Sinica, Taipei 100, Taiwan, ROC.*

Magnetic field dependence of low temperature specific heat of spinel oxide superconductor $LiTi_2O_4$ has been elaborately investigated. In the normal state, the obtained electronic coefficient of specific heat $\tilde{\gamma}_n$ = 19.15 mJ/mol $K^2$, the Debye temperature $\Theta_D$ = 657 K and some other parameters are compared with those reported earlier. The superconducting transition at $T_c \sim$ 11.4 K is very sharp ($\Delta T_c \sim$ 0.3 K) and the estimated $\Delta C/\tilde{\gamma}_n T_c$ is $\sim$ 1.78. In the superconducting state, the best fit of data leads to the electronic specific heat $C_{es}/\tilde{\gamma}_n T_c$ = 9.87 exp (-1.58 $T_c/T$) without field and $\tilde{\gamma}(H) \propto H^{0.95}$ with fields. In addition, $H_{c2}(0) \sim$ 11.7 T, $H_c(0) \sim$ 0.32 T, $\xi_{GL}(0) \sim$ 55 Å, $\lambda_{GL}(0) \sim$ 1600 Å, and $H_{c1}(0) \sim$ 26 mT are estimated from Werthamer-Helfand-Hohenberg (WHH) theory or other relevant relations. All results from the present study indicate that $LiTi_2O_4$ can be well described by a typical type-II, BCS-like, moderate coupling, and fully gapped superconductor in the dirty limit. It is further suggested that $LiTi_2O_4$ is a moderately electron-electron correlated system.

## I. INTRODUCTION

$LiTi_2O_4$ is unique among oxide superconductors in many respects like its chemistry, crystal structure and superconducting properties.[1-11] In the normal spinel-like structure (space group $Fd3m$) of $LiTi_2O_4$, the Li and Ti atoms are respectively at the positions of tetrahedral (8a) and octahedral (16d) sites. The resistivity and magnetic susceptibility data[1] of $LiTi_2O_4$ showed the $T_c \sim$ 10-12 K. The disappearance of superconductivity in $Li_{1+x}Ti_{2-x}O_4$ for x > 0.15 was concluded to be due to grain boundary effects.[1,7] It has attracted a lot of attentions due to at least having the following physical significances related to the present studies. For example, it is the only spinel oxide superconductor ($T_c \sim$ 12 K) so far to our knowledge. Also, it is the rare oxide superconductor showing a sharp superconducting anomaly[2-4] in specific heat ($C$) in contrast to an unpronounced one in polycrystalline $BaPb_{1-x}Bi_xO_3$ with a comparable $T_c \sim$ 12 K.[12] The upper critical field $H_{c2}(0)$ of $LiTi_2O_4$ reported by several groups varied from 2 to 32.8 T.[3,7] Issues like whether the superconductivity in $LiTi_2O_4$ can be well explained in the framework of BCS theory based upon the electron-phonon ($e$-$ph$) interactions and the role of the electron-electron ($e$-$e$) interactions have not been totally clearified.[9] Some theoretical predictions showed that $LiTi_2O_4$ was a strong coupling BCS superconductor while the low temperature specific heat and magnetic susceptibility data implied for the conditions for weak coupling $d$-band superconductivity.[1-4,10] Furthermore, there has been a very recently revived debate on Anderson's resonating valance bond (RVB)-type ground state as the possible origin of superconductivity in cuprates.[13,14] Since the Ti sub-lattice of the spinel structure allows a high degree of frustration, RVB ground state is probable in the $LiTi_2O_4$ spinel.[10]

In fact, the specific heat ($C$), a thermodynamic bulk property unlike resistivity and magnetization, of $LiTi_2O_4$ has been elaborately studied by some groups[2-4] in the absence of magnetic field ($H$). Though some of the derived parameters (listed in Table I) agree quite well with each other, some of them differ significantly and lead to incompatible descriptions for the nature of superconductivity. These controversies especially warrant a comprehensive revisit of superconductivity in $LiTi_2O_4$. In addition, the recent intensive investigations of its isostructural compounds $LiMn_2O_4$ (Ref. 15) and $LiV_2O_4$ (Refs. 4 and 16) exhibiting respectively high-voltage electrolyte and 3$d$ heavy fermion behavior also tempt us to study the crucial role of 3$d$ metals in the spinel oxide structure. In particular, it is interesting to study the evolution from 3$d$ superconductivity to 3$d$ heavy fermion in $Li(Ti_{1-x}V_x)_2O_4$.[17] In this report, we thus provide the detailed magnetic field (up to 8 T) dependence of the low temperature specific heat on $LiTi_2O_4$, which has never been reported in the literature earlier and is crucial to the determination of superconducting properties, to explore its pairing mechanism of superconductivity.

## II. EXPERIMENTAL METHODS

The preparation and characterization of polycrystalline $LiTi_2O_4$ used in this low temperature specific heat measurement were described elsewhere.[8] Highly pure $Li_2CO_3$ and $TiO_2$ were mixed in an appropriate ratio, calcined in a quartz tube under pure oxygen atmosphere for 20 h at 750 $^0$C leading to the formation of $Li_2Ti_2O_5$. Then it was mixed with a proper amount of $Ti_2O_3$, grounded thoroughly, pressed into pellets, and sintered at 880 $^0$C for 24 h under a dynamic vacuum with pressure less than $10^{-4}$ torr. Basically, to obtain a pure $LiTi_2O_4$ phase, one needs to add $\sim$ 15% more of Li than nominal composition due to its volatility. Powder x-ray diffraction (XRD) data obtained by SIEMENS D5000 diffractometer using $CuK\alpha$ radiation showed that $Li_{1+x}Ti_2O_4$ exhibited $Ti_2O_3$ impurity phase for x=0 and pure $LiTi_2O_4$ phase was obtained for $0.1 \leq x \leq 0.15$ (Ref. 8) which was used for this specific heat measurement. The low temperature specific heat $C(T,H)$ was measured with a $^3$He heat-pulsed thermal relaxation calorimeter[18] in the temperature range from 0.6 to 20 K under different magnetic fields (0 – 8 T). The precision of the measurement in this temperature range is about 1%. To test the accuracy of the field dependence of specific heat, $C(T,H)$ of a standard copper sample was measured at $H$ = 0, 1, and 8 T, respectively. The scatter of data in different magnetic fields was within 3%. $T_c$ (~ 11.4 K) obtained from specific heat data is consistent with that measured by resistivity on the same sample.[8]

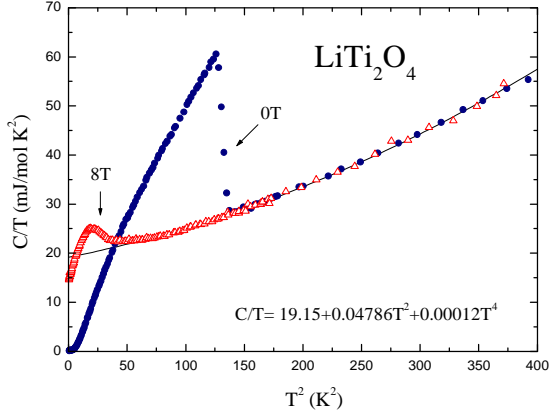

**FIG. 1.** $C(T,H)/T$ vs. $T^2$ of $LiTi_2O_4$ without and with a magnetic field of 8 T. The solid line is the best fit of $C_n(T)/T = \gamma_n + sT^2 + rT^4$ to the $H$ = 8 T data between 8 and 20 K.

## III. RESULTS AND DISCUSSION

Figure 1 shows the low temperature specific heat $C(T,H)$ of $LiTi_2O_4$ with $H$ = 0 and 8 T as $C/T$ vs. $T^2$. The normal state specific heat in the absence of magnetic field,

$$C_n(T) = \gamma_n T + C_{lattice}(T), \quad (1)$$

is extracted from $H$ = 8 T data between 8 and 20 K, where $\gamma_n T$ is the electronic term due to free charge carriers and $C_{lattice}(T) = sT^3 + rT^5$ is representing the phonon contribution which is assumed to be independent of magnetic field. It is found that $\gamma_n$ = 19.15 ± 0.20 (mJ/mol $K^2$), $s$ = 0.048 ± 0.002 (mJ/mol $K^4$), and $r$ = 0.00012 ± 0.00005 (mJ/mol $K^6$) give the best fitting (solid line in Fig. 1) to the experimental data. It is noted that if we take $C_{lattice}(T) = sT^3 + rT^5 + DT^7$, the best fit occurs at negative value of D ( = -2.8±0.5 x $10^{-7}$ mJ/mol $K^8$) which is unreasonable. On the other hand, $C_{lattice}(T) = sT^3$ gives much higher RMS value compared to $C_{lattice}(T) = sT^3 + rT^5$. The enhancement of $\gamma_n$ by the electron-phonon interaction is given by[19]

$$\gamma_n = (1/3)k_B\pi^2 N(E_F)(1+\lambda), \quad (2)$$

where $N(E_F)$ is the band structure density of states at the Fermi level, $k_B$ is the Boltzman constant, and $\lambda$ is the electron-phonon interaction constant. Taking $\lambda$ = 0.65 as obtained from low temperature specific heat data (discussed later), the calculated value of $N(E_F)$ for the present sample is ~ 0.70 states/eV atom. This is lower than that (~ 0.97 states/eV atom) perceived by McCallum et al.[2] from susceptibility data which would not satisfy the transition temperature from McMillan equation.[19] However, the authors[2] indicated that the reduction of $N(E_F)$ by 15% (orbital contribution and/or exchange enhancement to total susceptibility) would explain it. The reduced $N(E_F)$ is in close agreement with the present investigation. The Debye temperature $\Theta_D$ = 657 ±33 K is derived by using the relation

$$s = 1.944 \times 10^6 \times n/\Theta_D^3, \quad (3)$$

where $n$ is the number of atoms per formula unit and takes 7 for $LiTi_2O_4$. This is somewhat higher than those experimentally obtained and theoretically predicted values[3,11] (~575 K) but closer to those (685-700 K) reported by the group of D. C. Johnston.[2,4] The values of $\gamma_n$, $N(E_F)$, and $\Theta_D$ are listed in Table É along with some other parameters for comparison with reported results.

The characteristics of superconducting phase transition in $LiTi_2O_4$ can be analyzed using the relation

$$\delta C(T) = C(T, H = 0 \text{ T}) - C_n(T). \quad (4)$$

The resultant $\delta C(T)/T$ vs. T is shown in Fig. 2, where the inset illustrates the conservation of entropy $S = \int_0^{T_c} \frac{\delta C}{T} dT$ around the transition. This conservation of entropy is essential for a second order, such as superconducting-normal, phase transition. In this case, fail to include $rT^5$ term in Eq. (1) will not totally satisfy this requirement.[2] In fact, this requirement may be used to testify the justification of the values of $\gamma_n$ and $\Theta_D$. The dimensionless specific-heat jump at $T_c$ is $\delta C/\gamma_n T_c = 1.78$ as indicated in Fig. 2 which is greater than the typical weak coupling value (~1.43). Thus $\delta C(T)/T$ is well fitted to the BCS model as shown in Fig. 2 by the solid line with a little higher $2\Delta/k_BT_c$ (~ 4.0), where $\Delta$ is the superconducting energy gap instead of the weak coupling value (~ 3.52). This value of $2\Delta/k_BT_c \sim 4$ (i.e. $\Delta$ = 1.97 meV) is consistent with 3.8 in the Ref. [3] and the reference therein where the tunneling experiments yielded the value of 4.0. Consequently, *the superconductivity in $LiTi_2O_4$ can be explained by the moderate coupling BCS framework though the early low temperature specific heat and theoretical calculations respectively indicated weak and strong coupling.*[2-4,10-11]

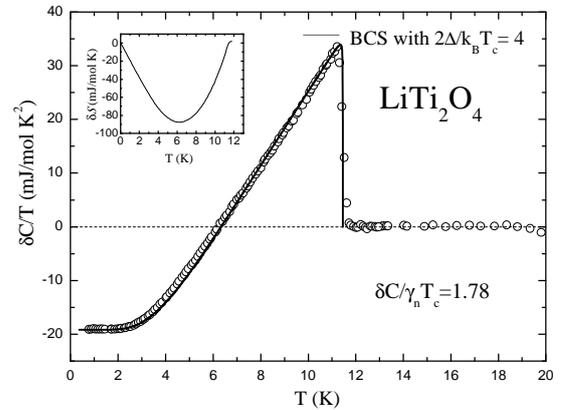

**FIG. 2.** Plot of $\delta C(T)/T$ vs. T with $\delta C(T) = C(T) – C_n(T)$. The solid line is the BCS fitting with $2\Delta/k_BT_c = 4$. The inset shows the entropy conservation around the transition temperature.

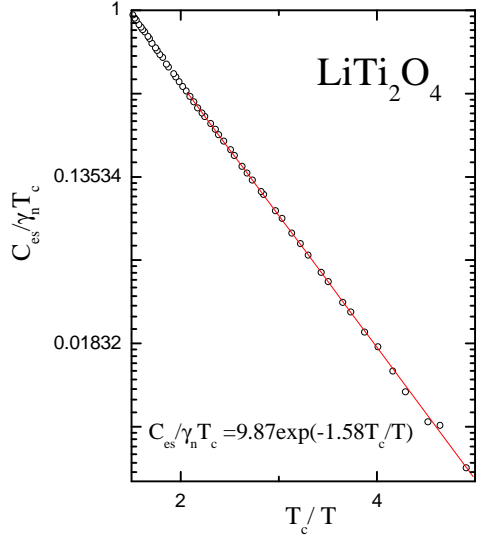

**FIG. 3.** Logarithmic $C_{es}/\gamma_n T_c$ vs. $T_c/T$ of LiTi$_2$O$_4$ in the superconducting state. The solid line is the linear fit to the data for $T_c/T$ between 2 and 5.

The electronic specific heat in the superconducting state is given by $C_{es}(T) = C(T) - C_{lattice}(T)$. Plot of logarithmic $C_{es}(T)/\gamma_n T_c$ vs. $T_c/T$ (Fig. 3) shows that the fitting of data (as demonstrated by the solid line) within $T_c/T$ = 2 to 5 and follows the relation $C_{es}(T)/\gamma_n T_c = A \exp[(-a\, T_c/T)]$ with $A$ = 9.87 and $a$ = 1.58. However, the BCS theory predicts $C_{es}(T)/\gamma_n T_c = 8.5 \exp[(-1.44\, T_c/T)]$ for this temperature fitting range in the weak coupling limit.[2] Therefore, both the values of coefficient and prefactor are in the range of typical moderate coupling BCS fully gapped superconductors.

The electron-phonon coupling constant $\lambda$ is estimated to be ~ 0.65 (Table I) using the relation[20]

$$\Delta C/\chi_n T_c = 1.43 + 0.942\lambda^2 - 0.195\lambda^3. \quad (5)$$

According to McMillan model,[19] for weak coupling $\lambda \ll 1$, for weak and intermediate coupling $\lambda < 1$, and for strong coupling $\lambda > 1$. Therefore, the present $\lambda$ value suggests that LiTi$_2$O$_4$ is a moderate coupling superconductor rather than a weak coupling one.[2-4] Nevertheless, the value of $\lambda$ is much lower than that (~ 1.8) of the theoretical predictions (indicating strong coupling superconductivity in LiTi$_2$O$_4$) which may be due to the spin fluctuation effect.[10,11] Taking $\Theta_D$ = 657 K, $\lambda$ = 0.65, and observed $T_c$ ~ 11.4 K, the Coulomb repulsion parameter $\mu^*$ ~ 0.13 can be obtained from the McMillan formula,[19]

$$T_c = (\Theta_D/1.45) \times \exp\{-1.04(1+\lambda)/[\lambda - \mu^*(1+0.62\lambda)]\}. \quad (6)$$

The value of $\mu^*$ is the same as transition metals,[19] and is in the range of those reported earlier[2-3] confirming the BCS-type $d$-band superconductivity in LiTi$_2$O$_4$.

The magnetic filed dependence of $C(T,H)/T$ and $\Delta C(T,H)(\equiv C(T,H) - C_n(T))/T$ were plotted in Fig. 4(a) and (b), respectively. It is noticed that the conservation of entropy (the area above and below zero of $C/T$ are equivalent around the superconducting transition) is fundamentally satisfied for all studied magnetic fields.

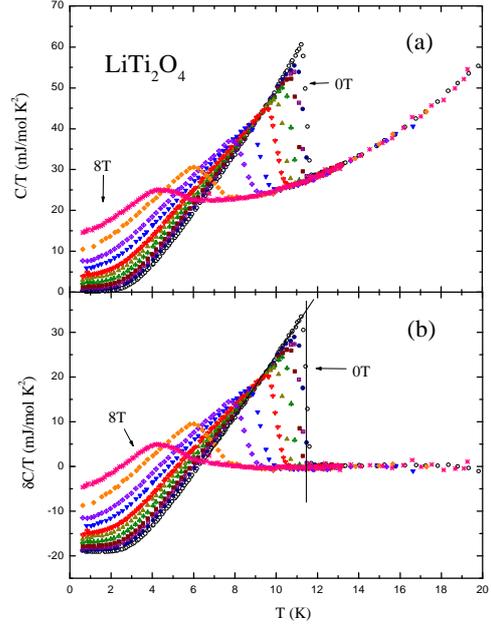

**FIG. 4.** (a) $C(T,H)/T$ vs. $T$ and (b) $\Delta C(T,H)/T$ vs. $T$ under various magnetic fields. The entropy around the superconducting transition is conserved even in magnetic fields.

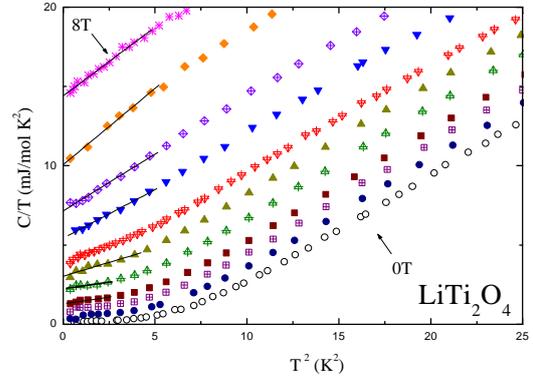

**FIG. 5.** $C/T$ vs. $T^2$ at very low temperature ($\leq$ 5 K) under different magnetic fields. $\gamma(H)$ has been estimated from the linear extrapolation of data down to 0 K.

This implies that the sample is of good quality without detectable impurities (particularly magnetic field dependent non-superconducting phases). The dissimilar example has been observed in Ba$_{1-x}$K$_x$BiO$_3$ (Ref. 21) due to defects and the inhomogeneity of the sample. Figure 5 demonstrates the magnetic field dependence of very low temperature ($\leq$ 5 K) specific heat as $C/T$ vs. $T^2$. The coefficient of electronic specific heat $\gamma(H)$ with various fields has been estimated from the linear extrapolation of data below 2 K down to 0 K. In order to investigate the low energy vortex excitation under magnetic fields, the variation of $\gamma(H)$ with $H$ is shown in Fig. 6. The best fit leads to $\gamma(H) \sim H^{0.95}$ as indicated by the solid line. Clearly, $\gamma(H)$ follows an $H$ dependence which is very close to be linear, especially for $H \geq 1$ T. The slight deviation in low

$H$ could be due to the vortex-vortex interaction as discussed in Refs. 22 and 23. The value of $H_{c2}(T=0)$ estimated from Fig. 6 using linear extrapolation of $\gamma(H)$ for $H \geq 1$ T to $\tilde{a}_n \sim 19.15$ mJ/mol K$^2$ is $H_{c2}(T=0) = 11.0 \pm 0.5$ T. It is noted that a pronounced nonlinearity of $\gamma(H)$, seen in UPt$_3$,[24] CeRu$_2$,[25] and NbSe$_2$,[26] at low magnetic fields, is not obvious in the present LiTi$_2$O$_4$. Theoretically, $\gamma(H)$ is expected to be proportional to $H$ for a conventional $s$-wave superconductor.[27] However, $\gamma(H) \propto H^{0.5}$ is predicted for a nodal superconductivity.[28] In fact, $\gamma(H)$ of cuprate superconductors has been extensively studied in this context.[29] Consequently, the magnetic field dependence of $\gamma(H)$ suggests that LiTi$_2$O$_4$ is an $s$-wave superconductor in nature.

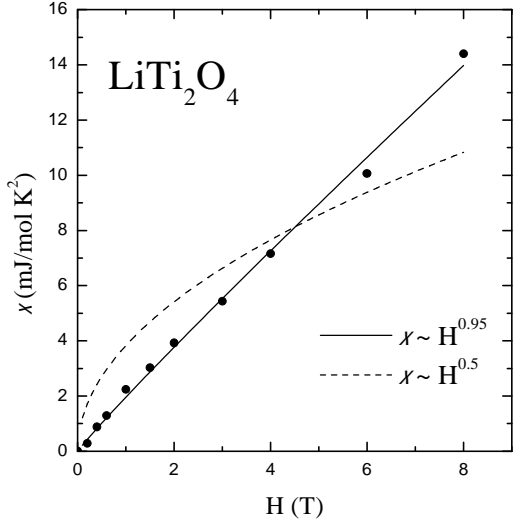

**FIG. 6.** Magnetic field dependence of electronic specific heat coefficient $\gamma(H)$ derived from the linear extrapolation of data using $C/T$ vs. $T^2$ plot (Fig. 5) for various magnetic fields below 2 K.

Figure 7 shows the temperature variation of upper critical field $H_{c2}(T)$ as obtained from Fig. 4 where the solid line is the theoretical fitting based on negligible spin paramagnetic and spin-orbital effect by using the Werthamer-Helfand-Hohenberg (WHH) theory.[30] The error bar in $T_c(H)$ is determined by the sharpness of the superconducting transition as shown in the inset of Fig. 7 for a particular magnetic field of 6 T. The same procedure is also followed for other magnetic fields. The best fit results in $(dH_{c2}/dT_c)_{T=Tc} = 1.45 \pm 0.03$ (T/K) and consequently $H_{c2}(0) = 11.7 \pm 0.4$ T (Table I). It is noted that the value of $H_{c2}(0)$ estimated from WHH theory is consistent with the value ($11.0 \pm 0.5$ T) obtained from Fig. 6. This consistency implies that the spin-orbital interaction in LiTi$_2$O$_4$ is negligible as considered for the fitting of the data with WHH theory. The small spin-orbital interaction is actually expected since Ti is one of the lightest transition elements. However, the $H_{c2}(0)$ of present sample is much higher than that ($\sim 2$ T) predicted by Heintz et al.[3] but lower than that ($\sim 32.8$ T) reported by Harrison et al.[7] For type-II superconductors, the Pauli limiting field $H_p(0) = 1.84 \times 10^4 T_c$ should satisfy the relation[31] $H_{c2}(0) \leq H_p(0)$. Though Harrison et al.[7] concluded LiTi$_2$O$_4$ to be an extreme type-II superconductor, their reported value of $H_{c2}(0)$ did not satisfy the above condition. Our estimated value of $H_p(0) \sim 21$ T (Table I) is higher than $H_{c2}(0) \sim 11.7$ T confirming the typical type-II superconductivity in LiTi$_2$O$_4$.

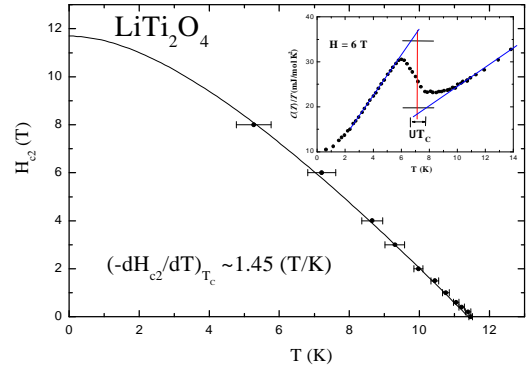

**FIG. 7.** Temperature variation of upper critical field $H_{c2}(T)$ of LiTi$_2$O$_4$ obtained from Fig. 4 where the solid line indicates the fitting of data with WHH theory by which $H_{c2}(0) \sim 11.7$ T is estimated. Inset shows the example how the error bar of $T_c(H)$ is determined by the sharpness of the superconducting transition for $H = 6$ T data.

To determine the following important parameters of LiTi$_2$O$_4$, the residual resistivity $\tilde{n}_{res} \sim 6.9 \times 10^{-5}$ Ω-cm was calculated from the formula[32]

$$\left[-dH_{c2}/dT\right]_{T=T_c} = 4.48 \times 10^4 \tilde{a}_n \tilde{n}_{res}. \qquad (7)$$

It is noted that the calculated $\tilde{n}_{res}$, though much lower than the measured value ($\sim 5.3 \times 10^{-2}$ Ω-cm) from the resistivity data (not shown), is in the same order of magnitude as that reported in Ref. 7. However, the measured value is consistent with that reported by Johnston et al.[1] This large discrepancy between the measured and calculated $\tilde{n}_{res}$ indicates that the studied sample exhibits a lot of grain boundaries which highly affect its electrical transport properties. The single crystalline LiTi$_2$O$_4$ may be indispensable for solving the puzzles of the transport properties. Thermodynamic critical field $H_c(0) \sim 0.320 \pm 0.003$ T, Ginzburg-London (GL) coherence length $\xi_{GL}(0) \sim 55 \pm 3$ Å, penetration depth $\lambda_{GL}(0) \sim 1600 \pm 50$ Å, and lower critical field $H_{c1}(0) \sim 26.0 \pm 0.3$ mT (Table I) are estimated from the following relations[32,33] in the dirty limit by using $\tilde{a}_n \sim 4570$ erg/cm$^3$K$^2$ (converted from $\tilde{a}_n \sim 19.15$ mJ/mol K$^2$), $\tilde{n}_{res} \sim 6.9 \times 10^{-5}$ Ω-cm (calculated from Eq. 7) and $H_{c2}(0) \sim 11.7$ T,

$$H_c(0) = 4.23 \tilde{a}_n^{1/2} T_c \text{ Oe}, \qquad (8)$$
$$\xi_{GL}(0) = \{\Phi_0/[2\pi H_{c2}(0)]\}^{1/2} \text{ Å}, \qquad (9)$$
$$\lambda_{GL}(0) = 6.42 \times 10^5 (\tilde{n}_{res}/T_c)^{1/2} \text{ Å}, \qquad (10)$$
$$H_{c1}(0) = H_c(0)(2^{1/2}\kappa)^{-1}\ln\kappa \text{ Oe}, \qquad (11)$$

where the fluxon $\Phi_0 \sim 2.0678 \times 10^9$ Oe Å$^2$ and $\kappa$ [$= \lambda_{GL}(0)/\xi_{GL}(0)$] $\sim 29 \pm 1$. The values of $\xi_{GL}(0)$ and $\lambda_{GL}(0)$ are, respectively, higher and lower than those ($\xi_{GL}(0) \sim 25.9$ Å and $\lambda_{GL}(0) \sim 2730$ Å) reported by Harrison et al.[7] Furthermore, $H_{c1}(0)$ is in the range of the values achieved from[3] the M - H curve ($\sim 20$ mT) and the fitting of $H_{c1}(T) = H_{c1}(0)(T_c-T)$ plot ($\sim 25$ mT). All the obtained parameters $\xi_{GL}(0)$, $\lambda_{GL}(0)$, and $\kappa$ of LiTi$_2$O$_4$ further satisfy the conditions for type-II superconductivity.[33]

All the above estimation of the parameters assumes the dirty limit superconductivity in LiTi$_2$O$_4$. Consequently, it is of interest to estimate the value of mean free path $l$ below $T_c$, which should be smaller than that of $\xi_{GL}(0)$ in the dirty limit. The band structure calculations indicate that the Fermi level of LiTi$_2$O$_4$ in the partially filled conduction band lies in an electronic structure which is not too far from the free-electron-like one with a mass renormalization factor.[10,11] By the free electron model with $\rho_{res} = 6.9\times10^{-5}$ Ω-cm and the carrier concentration $n = 1.35\times10^{23}$ cm$^{-3}$ (Ref. 7), the estimated $l = 32$ Å is indeed shorter than $\xi_{GL}(0)$. Therefore, the above analysis using Eq. (7)-(11) in the dirty limit regime is self-consistent. In addition, the issue of the mass renormalization factor in the previous literature[10,11] was actually unsolved. A large $\lambda_{tot} = 1.8$ was inferred implying a strong electron-electron interaction with unknown origin.[11] One can express $\lambda_{tot} = \lambda + \lambda_e$, where $\lambda$ is the electron-phonon coupling constant and $\lambda_e$ manifests the interactions due to the possible spin fluctuations and other electron-electron interactions. If we rewrite Eq. (2) as $\gamma_n = (1/3)k_B\pi^2 N(E_F)(1+\lambda_{tot})$ and assume the theoretical $N(E_F) = 0.46$ states/eV atom as achieved from the band structure calculations,[10,11] then $\lambda_{tot} = 1.53$ can be obtained from the $\gamma_n = 19.15$ (mJ/mol K$^2$) of present specific heat data. Therefore, the resultant $\lambda_e = \lambda_{tot} - \lambda = 1.53 - 0.65 = 0.88$ suggests a moderate electron-electron interaction in LiTi$_2$O$_4$, and is more consistent with a Stoner enhancement factor $(1-S)^{-1} \sim 2$, which was derived from the magnetic susceptibility $\chi = \mu_B^2 N(E_F)/(1-S)$.[1,11]

**TABLE I.** Some important parameters of LiTi$_2$O$_4$ obtained from the present and earlier investigations.

| Parameters | Ref. 2 | Ref. 3 | Ref. 4 | Present work |
|---|---|---|---|---|
| $T_c$ (K) | 11.7 | 12.4 | 11.8 | 11.4±0.3 |
| $\Delta T_c$ (K) | 1.2 | 0.32 | 0.2 | 0.3 |
| $\gamma_n$ (mJ/mol K$^2$) | 21.4 | 21.98 | 17.9 | 19.15±0.20 |
| $N(E_F)$ (states/eV atom) | 0.97 | 0.76 | 0.82 | 0.70±0.01 |
| $\beta$ (mJ/mol K$^4$) | 0.043 | 0.089 | 0.040 | 0.048±0.002 |
| $\Theta_D$ (K) | 685 | 535 | 700 | 657±33 |
| $\Delta C/\gamma_n T_c$ | 1.59 | 1.57 | 1.75 | 1.78 |
| $2\Delta/k_B T_c$ | -- | ~3.8 | -- | ~ 4.0 |
| $\Delta$ (meV) | -- | -- | -- | 1.97 |
| $\lambda$ | 0.64 | 0.71 | 0.63 | 0.65 |
| $H_{c2}(0)$ (T) | -- | >2 | -- | 11.0±0.5 |
| $H_c(0)$ (T) | -- | -- | -- | 0.327±0.003 |
| $H_{c1}(0)$ (mT) | -- | 20-25 | -- | 26.3±0.3 |
| $H_p(0)$ (T) | -- | -- | -- | 21.0±0.4 |
| $l$ (Å) | | | | 32 |
| $\xi_{GL}(0)$ (Å) | -- | -- | -- | 55±3 |
| $\lambda_{GL}(0)$ (Å) | -- | -- | -- | 1600±50 |
| $\kappa$ | -- | -- | -- | 29±1 |

**TABLE II.** Comparison of several important parameters for particular transition-metal oxide superconductors.

| Superconductors | LiTi$_2$O$_4$ | BaPb$_{0.75}$Bi$_{0.25}$O$_3$ | Ba$_{0.6}$K$_{0.4}$BiO$_3$ | La$_{1.84}$Sr$_{0.16}$CuO$_4$ | Sr$_2$RuO$_4$ |
|---|---|---|---|---|---|
| Crystal structure | Spinel cubic | Perovskite cubic | Perovskite cubic | Layered-perovskite | Layered-perovskite |
| $T_c$ (K) | 11.4±0.3 | 11.7 | 30 | 38 | 1.48 |
| $\gamma_n$ (mJ/mol K$^2$) | 19.15±0.20 | 1.6 | 0.9 | 0.77 | 37.5 |
| $N(E_F)$ (states/eV atom) | 0.70±0.01 | 0.14 | 0.32±0.07 | -- | -- |
| $\Theta_D$ (K) | 657±33 | 195 | 346 | 389 | -- |
| $\Delta C/\gamma_n T_c$ | 1.97 | -- | -- | -- | 0.74±0.02 |
| $2\Delta/k_B T_c$ | ~ 4 | -- | 3.5±0.1 | -- | -- |
| $\lambda$ | 0.65 | 1.45 | 0.6-0.8 | -- | -- |
| Pairing state | $s$-wave | $s$-wave | $s$-wave | $d$-wave | $p$-wave ? |
| Reference | This work | 12 | 21,34 | 35,36 | 37 |

Finally, it may be interesting to look over the existing transition-metal oxide superconductors, such as, BaPb$_{0.75}$Bi$_{0.25}$O$_3$ (Ref. 12), Ba$_{0.6}$K$_{0.4}$BiO$_3$ (Refs. 21 and 34), La$_{1.84}$Sr$_{0.16}$CuO$_4$ (Refs. 35 and 36) and Sr$_2$RuO$_4$ (Ref. 37) for comparison. Some of the important parameters along with those of our studied LiTi$_2$O$_4$ are summarized in Table II. One would find that the $T_c$ of these superconductors does not strongly correlate with their structure, $\gamma_n$, $N(E_F)$, $\Theta_D$, and the electron-phonon coupling constant $\lambda$. It is also evident that the superconductivity of each material occurs at only very narrow transition-metal composition. A small amount of metal substitution or a little off-stoichiometry for transition metal will dramatically suppress the superconductivity. Thus, the electronic properties of the transition metals Ti, Bi, Cu, and Ru must play an unique role on the occurrence of superconductivity. In addition, the superconducting pairing state of these superconductors varies from $s$-wave, $d$-wave to possible $p$-wave symmetry (Table II). Accordingly, theoretical calculations and experimental probes on the energy bands need to be done for clarifying these points. Moreover, three more transition-metal oxide superconductors Cd$_2$Re$_2$O$_7$ ($T_c \sim 1$ K, Ref. 38), Na$_{0.35}$CoO$_2$ · 1.3H$_2$O ($T_c \sim 5$ K, Ref. 39) and KOs$_2$O$_6$ ($T_c \sim 9$ K, Ref. 40) were reported most recently and also have attracted much attention due to their novel normal- and superconducting-state features. It is no doubt that the understanding of nature of superconductivity in transition-metal oxide superconductors will still challenge the scientists in the fields of condensed matter physics.

## IV. SUMMARY

In summary, the low temperature specific heat of LiTi$_2$O$_4$ in magnetic fields is presented. Based on the present measurements and relevant theoretical relations, the normal- and superconducting-state parameters including electronic specific heat coefficient $\gamma_n = 19.15$ mJ/mol K$^2$, Debye temperature $\Theta_D = 657$ K, $\Delta C/\gamma_n T_c \sim 1.78$, superconducting energy gap $\Delta \sim 1.97$ meV, electron-phonon coupling constant $\lambda \sim 0.65$, upper critical field $H_{c2}(0) \sim 11.7$ T, thermodynamic critical field $H_c(0) \sim 0.32$ T, coherence length $\xi_{GL}(0) \sim 55$ Å, penetration depth $\lambda_{GL}(0) \sim 1600$ Å, and lower critical field $H_{c1}(0) \sim 26$ mT are evaluated and compared with some of those reported. Combining the results $C_{es}(T)/\gamma_n T_c = 9.87\exp(-1.58T_c/T)$ and $\chi(H) \propto H^{0.95}$, we conclude that LiTi$_2$O$_4$ is a typical BCS-like, fully gapped, and moderate-coupling type-II superconductor in the dirty limit. The analysis also suggests that LiTi$_2$O$_4$ is a moderately electron-electron correlated system.

## ACKNOWLEDGEMENT
This work was supported by National Science Council of Republic of China under contract Nos. NSC 92-2112-M-110-017 and NSC 92-2112-M-009-032.

*Corresponding author: yang@mail.phys.nsysu.edu.tw
†Permanent address: Department of Physics, Aligarh Muslim University, Aligarh-202002, India.